\documentclass[twocolumn,showpacs,floatfix]{revtex4}
 
\usepackage{graphicx}  \usepackage{amsmath}

\begin{document}

\title{Quantum gates between capacitively coupled double quantum dot
two-spin qubits}

\author{Dimitrije Stepanenko} \author{Guido Burkard}
\affiliation{Department of Physics and Astronomy, University of Basel,
Klingelbergstrasse 82, CH-4056 Basel, Switzerland}

\begin{abstract} We study the two-qubit controlled-not gate operating
    on qubits encoded in the spin state of a pair of electrons in a
    double quantum dot.  We assume that the electrons can tunnel
    between the two quantum dots encoding a single qubit, while
    tunneling between the quantum dots that belong to different qubits
    is forbidden.  Therefore, the two qubits interact exclusively
    through the direct Coulomb repulsion of the electrons.  We find
    that entangling two-qubit gates can be performed by the electrical
    biasing of quantum dots and/or tuning of the tunneling matrix
    elements between the quantum dots within the qubits.  The
    entangling interaction can be controlled by tuning the bias
    through the resonance between the singly-occupied and
    doubly-occupied singlet ground states of a double quantum dot.
\end{abstract}

\pacs{73.21.La,03.67.Lx,85.35.Be}

\maketitle

\section{Introduction}

The spin-1/2 of a single electron trapped in a quantum dot (QD) is a
promising candidate for a carrier of quantum information in a quantum
computer \cite{LD98}.  To perform a quantum computation we need to
have all the unitary operations from some universal set of quantum
gates at our disposal \cite{NC00}.  One such universal set consists of
all the single qubit quantum gates and a two-qubit controlled-not
(CNOT) quantum gate.  Quantum computation over the single-spin qubits
with the logical states corresponding to the spin orientations
$|\!\!\uparrow\rangle$ and $|\!\!\downarrow\rangle$ can in principle
be achieved using an external magnetic field or with g-factor
engineering for the single qubit operations, and with the
time-dependent isotropic exchange interaction $H_{ex}(t)=J(t){\bf
S}_1\cdot {\bf S}_2$ for manipulating a pair of qubits encoded into
spins ${\bf S}_1$ and ${\bf S}_2$ \cite{LD98}.

Control of electron spins in quantum dots is in the focus of many
intense experimental investigations.  Manipulation of pairs of
electron spins using the tunable isotropic exchange interaction has
already been demonstrated in several experiments
\cite{PJT+05,KFE+05,JPT+05}.  Such control was used in a study of the
QD spin decoherence due to the hyperfine coupling to the surrounding
nuclear spins, where the splitting between the singlet states with
the total spin $S=0$ and the triplet states with $S=1$ was used to
turn on and off the singlet-triplet mixing caused by the hyperfine
interaction.  An important result of these studies is that the
coherence time of an electron spin in a quantum dot is very long if
the decoherence due to the interaction with the nuclear spins can be
suppressed.  The spin coherence times can be improved by the
manipulation of nuclear spins \cite{CL05,SBG+06,KCL06}, in principle
allowing for elaborate sequences of operations to be performed.
Single spin control is based on the local manipulation of the magnetic
field or g-factor \cite{LD98}, or on ESR methods \cite{BLD99,EL01} and
has only recently been demonstrated experimentally \cite{KBT+06}.

The difficulty of single-spin control has inspired a number of
proposals for quantum computation based on the encoding of qubits into
more than one spin.  These encoding schemes reduce the requirement on
the control over electron spins, but have the drawback of introducing
so-called leakage errors in which the state of encoded qubit ``leaks''
out of the set of computational states.  Standard error-correction
procedures can be modified to prevent this kind of errors \cite{AT07}.
A universal set of quantum gates operating on qubits encoded into
states of three quantum dot spins with equal total spin quantum
numbers can be implemented through control of the isotropic exchange
coupling alone $H_{ex}$ \cite{BKL+00,KBL+01,DBK+00}.  Control over
interactions that are symmetric only with respect to rotations about a
fixed axis in spin space allows for the construction of a universal
set of quantum gates that operate over qubits encoded into pair of
spins.  One such encoding is into the orthogonal  states $|\!\!
\uparrow \downarrow \rangle$ and $|\!\! \downarrow \uparrow\rangle$ of
two spins-1/2.  A universal set of quantum gates over such qubits can
in principle be performed by the control over $H_{ex}$, with the
anisotropy provided by an external static homogeneous magnetic field
and a site-dependent $g$-factor \cite{L02,B01}.

We consider a variant of the two-spin encoding where the logical zero
$|0_{\rm L}\rangle$ and the logical one $|1_{\rm L}\rangle$ quantum
states are the singlet and the triplet with zero projection of the
total spin to the symmetry axis $z$ ($S_z=0$), e.g., for lateral QDs,
the $z$ axis is the normal to the plane of the heterostructure,
\begin{equation}
\label{eq:stqubit}
\begin{split}
|0_L\rangle &=  \frac{1}{\sqrt{2}}\left(|\!\uparrow \downarrow\rangle
- |\!\downarrow \uparrow \rangle \right), \\ |1_L\rangle &=
\frac{1}{\sqrt{2}}\left(|\!\uparrow \downarrow\rangle + |\!\downarrow
\uparrow \rangle \right).
\end{split}
\end{equation}
These qubits can be manipulated by an axially symmetric interaction to
produce a universal set of quantum gates.  The interaction with an
external magnetic field and the isotropic exchange \cite{L02,B01}, or
the interaction with an external magnetic field and an anisotropic
spin-orbit coupling \cite{WL02}, or the spin-orbit coupling alone
\cite{SB04}, were all proposed as a way of producing a universal set
of quantum gates operating on singlet-triplet two-spin qubit,
Eq.~(\ref{eq:stqubit}).  Recently, it was suggested that an
architecture based on singlet-triplet qubits individually addressed
using isotropic exchange interaction and inhomogeneous magnetic field
and coupled through Coulomb interaction of the electrons is scalable
and in principle realizable \cite{TED+05}.

\begin{figure}
\includegraphics[width=8.5cm]{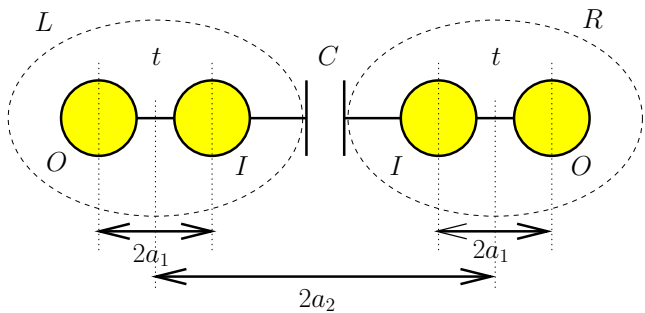}
\caption{Double-double quantum dot (DDQD) setup.  The four
  single-electron quantum dots are aligned along a fixed direction.
  The spins of the electrons on two quantum dots, inner (I) and outer
  (O), separated by a distance $2a_1$ encode a qubit.  Two such double
  quantum dot (DQD) qubits, left (L) and right (R), at the distance
  $2a_2$ are separated by an impenetrable barrier.  The tunneling
  matrix element $t$ within the double quantum dots (DQD) carrying
  the qubits, and the bias $\epsilon$ of the inner dots with respect
  to the outer are equal on both DQDs, and can be electrically tuned.
  The Coulomb interaction between the DQD is represented by the
  capacitor ${\rm C}$.\label{fig:4dots}}
\end{figure}

In this paper, we study a particular realization of entangling
two-qubit gates between singlet-triplet qubits,
Eq.~(\ref{eq:stqubit}), where each qubit is represented by a pair of
tunnel-coupled single-electron quantum dots, as proposed in
\cite{TED+05}.  In this realization, the double quantum dots are
separated by a barrier which is impenetrable for the electrons, so
that the qubits are coupled exclusively through the Coulomb repulsion
of electrons, while the exchange terms between electrons on different
double quantum dots vanish.  The setup of this double-double quantum
dot (DDQD) is illustrated in Fig.~\ref{fig:4dots}.

The Coulomb interaction is spin-independent, leading to an isotropic
interaction $J{\bf S}_1\cdot{\bf S}_2$ between tunnel-coupled spins
${\bf S}_1$ and ${\bf S}_2$.  The anisotropic correction to this
interaction is dominated by the spin-orbit coupling induced term
$J\mbox{\boldmath{$\beta$}}\cdot \left({\bf S}_1 \times {\bf
S}_2\right) + O(|\mbox{\boldmath{$\beta$}}|^2)$.  The relative
strength of the anisotropic interaction in the quantum dot systems in
GaAs is estimated to be $|\mbox{\boldmath{$\beta$}}|\sim0.1-0.01$
\cite{K01,K04}.  The influence of the anisotropic corrections can be
reduced in specific implementations of the quantum gates
\cite{BL02,BSD01}.  In our study of a two-qubit gate operation, we
will only consider the case of isotropic interaction and neglect the
weak anisotropy.  In this case, transitions between spin-singlet and
spin-triplet states on a DQD are forbidden.  Due to this spin
symmetry, the four-electron Hamiltonian is block-diagonal,
\begin{equation}
\label{eq:hblocks}
H={\rm diag}\left(H_{SS},H_{ST},H_{TS},H_{TT}\right).
\end{equation}
The non-zero blocks $H_{ab}$, where $a,b=S,T$, act on the states in
which electron pairs on each DQD are either in the singlet ($S$) or in
a triplet ($T$) state of the total spin $S=0$ or $S=1$.

Our main results are the effective low-energy spin interaction and a
scheme to perform a two-qubit CNOT gate in an electrically controlled
DDQD system.  The effective low-energy spin interaction in this setup
has the form
\begin{equation}
\label{eq:hstrongbias}
H=J\left({\bf S}_{LI}\cdot {\bf S}_{LO}+{\bf S}_{RI}\cdot {\bf
  S}_{RO}\right) + E_{e}|SS\rangle \langle SS|.
\end{equation}
Two pairs of spins, ${\bf S}_{LI}$ and ${\bf S}_{LO}$ on the left
$(L)$ qubit and ${\bf S}_{RI}$ and ${\bf S}_{RO}$ on the right $(R)$
qubit (see Fig.~\ref{fig:4dots}) interact via the isotropic exchange
interaction of strength $J$, and the entangling interaction of
strength $E_{e}$ that shifts the energy of the singlet-singlet state.
We show how the entangling two-qubit quantum gates for universal
quantum computation can be performed through the electrical control of
$E_{e}$.

The triplet states with $S_z=0,\pm 1$ are degenerate in the absence of
a magnetic field.  A uniform magnetic field ${\bf B}$, pointing along
the $z$ axis  normal to the plane of QDs causes a Zeeman splitting
$g\mu_B {\bf B}\cdot {\bf S}$ between the $S_z=0$ states and the
states with $S=1$, $S_z=\pm 1$.  Our results apply both to the
isotropic (${\bf B}=0$) and anisotropic, but axially symmetric (${\bf
B}\neq 0$) case, if we take the $S_z=0$ state to represent the qubit
$|1_L\rangle$ state.

\begin{figure}[t]
\includegraphics[width=8.5cm]{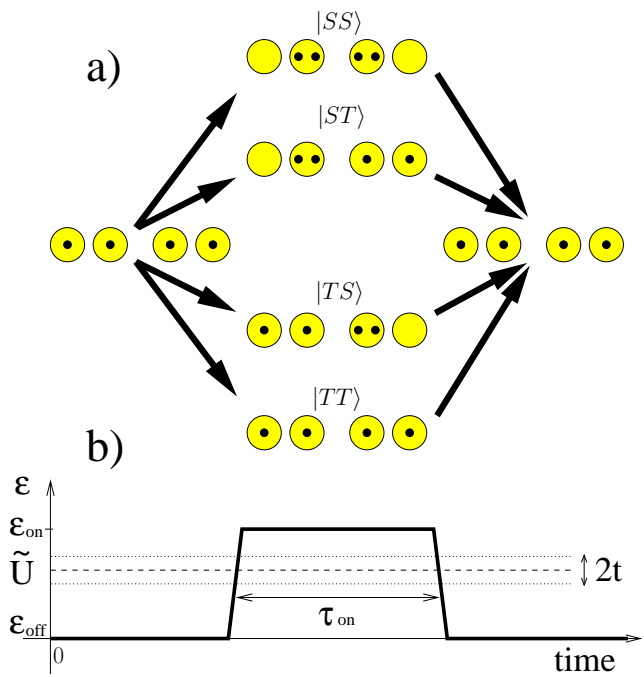}
\caption{Two-qubit quantum gate. a) When the inner quantum dots of the
  two double quantum dots system are strongly biased ($\epsilon >
  {\tilde U} + t$) the ground state is the doubly occupied inner dot.
  Due to the Pauli principle, only the spin singlets $(S)$ can tunnel
  into the doubly occupied states on their DQDs.  As the bias
  $\epsilon$ is reduced, the states again become degenerate.  b) A
  quantum gate is performed by sending a bias pulse $\epsilon(t')$.
  Each qubit state $|ab\rangle$ acquires a phase
  $\phi_{ab}=\int_{-\infty}^{\infty}E_{ab}(t')dt'/\hbar$, where
  $E_{ab}(t')$ is the ground state energy of the Hamiltonian at time
  $t'$ reduced to the appropriate spin subspace, resulting in a
  two-qubit quantum gate.
\label{fig:dotpulse}}
\end{figure}

A two-qubit quantum gate can in principle be performed by
adiabatically varying the tunneling matrix element $t$ and the bias
$\epsilon$ within the DQD.  In practice, it is much simpler to change
the bias $\epsilon$ while $t$ remains fixed \cite{HB06}.  The control
parameters $\epsilon$ and $t$ have to vary slowly on the time scale
set by the energy splitting between the states of a given spin
configuration.  During the gate application, the orbital components of
the $S$ and $T$ states are different due to the Pauli principle that
forbids the electrons in a spin triplet to share their orbital state,
see Fig.~\ref{fig:dotpulse}.
As opposed to $t$ and $\epsilon$ that are determined by gate voltages
and can be changed more or less at will, the Coulomb interaction is
set by the geometry of the system and therefore fixed.  We show how
the control of the parameters $t$ and $\epsilon$, or even $\epsilon$
alone, can nevertheless be used to implement entangling two-qubit
gates on encoded singlet-triplet qubits through its influence on the
Coulomb terms.

When an adiabatic gate is applied, the lowest energy state in each
block $H_{ab}$, of energy $E_{ab}$, where $a,b=S,T$, see
Eq.~(\ref{eq:hblocks}) acquires a phase
$\phi_{ab}=\int_{t_i}^{t_f}E_{ab}(t')dt'/\hbar$.  The energy $E_{ab}$
becomes time-dependent through the time-dependence of the parameters
$t$ and $\epsilon$ in the interval $t_i<t'<t_f$.  The resulting
interaction is described by an effective 4-dimensional two-qubit
Hamiltonian acting in the space spanned by the lowest-energy states
$|SS\rangle$, $|ST\rangle$, $|TS\rangle$, and $|TT\rangle$ in the
corresponding blocks $H_{ab}$, and has the form of
Eq.~(\ref{eq:hstrongbias}).

In the regime of strong bias, $|\epsilon -  U|\gg t$, where $U$ is the
on-site Coulomb repulsion, we investigate the DDQD system using
perturbation theory.  For the case of arbitrary bias $\epsilon$, we
numerically diagonalize the Hamiltonian Eq.~(\ref{eq:hblocks}).  We
show that the two-qubit quantum gate can be operated by tuning the
bias $\epsilon$ so that the amplitude of the doubly occupied state in
the lowest energy spin singlet becomes appreciable.  In this ``on''
state with a large double occupancy amplitude, entanglement is
generated between the two-spin qubits.  The entanglement generation is
suppressed in the ``off'' regime with weak bias and tunneling.
Therefore, the generation of entanglement between the two two-spin
qubits encoded into DDQD can be efficiently controlled using the bias
$\epsilon$ alone.  Together with the single-qubit operation this
control is sufficient for universal quantum computing.

This paper is organized as follows.  In Sec.~II, we introduce our
model of the DDQD system, followed by the discussion of the control
through voltage pulses.  In Sec.~III we focus on the case of the
strongly biased ($|\epsilon - U| \gg t$) DDQD system and calculate the
interaction between the qubits.  The constraint of the strong bias is
lifted in Sec.~IV, where we numerically find the interaction between
the qubits, valid at an arbitrary bias $\epsilon$.  In Sec.~V, we
outline the construction of a CNOT gate based on the resources for the
control over a pair of qubits deduced from the results of Secs.~III
and IV.  Our results are summarized in Sec.~VI. The technical details
of the calculation are collected in the Appendix A.

\section{Model}

For the purpose of finding the effective low-energy spin Hamiltonian,
the excited orbital states of single quantum dots can be neglected,
leading to the Hund-Mulliken (HM) approximation with one orbital per
dot \cite{BLD99,HD00}.  In the HM approximation, the state space of
the two-electron system in a double quantum dot (DQD) encoding the
left ($q=L$) or the right ($q=R$) qubit is spanned by three singlet
basis states, $|{\bar S}\rangle$, $|D_I\rangle$ and $|D_O\rangle$ and
one triplet basis state $|T_0\rangle$
\begin{align}
\label{eq:states}
|{\bar S}\rangle&=\frac{1}{\sqrt{2}}\left(
 c_{qI\uparrow}^{\dagger}c_{qO\downarrow}^{\dagger} -
 c_{qI\downarrow}^{\dagger}c_{qO\uparrow}^{\dagger}\right)|0\rangle,
 \\
 |D_I\rangle&=c_{qI\uparrow}^{\dagger}c_{qI\downarrow}^{\dagger}|0\rangle,
 \\
 |D_O\rangle&=c_{qO\uparrow}^{\dagger}c_{qO\downarrow}^{\dagger}|0\rangle,
 \\
\label{eq:statesend}
|T_0\rangle&=\frac{1}{\sqrt{2}}\left(
 c_{qI\uparrow}^{\dagger}c_{qO\downarrow}^{\dagger} +
 c_{qI\downarrow}^{\dagger}c_{qO\uparrow}^{\dagger}\right)|0\rangle,
\end{align}
where $c_k$ is the annihilation operator for an electron in the state
$k=(q_k,p_k,s_k)$ on the qubit $q_k=L,R$, with position $p_k=I,O$,
where $I$ stands for inner and $O$ for outer quantum dot within a
qubit, and spin $s_k=\uparrow,\downarrow$.  The vacuum $|0\rangle$ is
the state of empty QDs.

In the standard notation the singlet states of a DQD are denoted by
$|(n,m)S\rangle$, where $n$ is the number of electrons on the left QD
and $m$ is the number of electrons on the right QD.  Our singly
occupied singlet is then expressed as $|{\bar S}\rangle\equiv
|(1,1)S\rangle$.  The doubly occupied singlet states on the left,
$q=L$, DQD are $|D_I\rangle\equiv |(0,2)S\rangle$, and
$|D_O\rangle\equiv |(2,0)S\rangle$.  On the right, $q=R$, DQD the
definitions are reversed,  $|D_I\rangle\equiv |(2,0)S\rangle$, and
$|D_O\rangle\equiv |(0,2)S\rangle$.

The orbital states annihilated by $c_k$ approximate the ground states
of the single-particle Hamiltonian
\begin{equation}
\label{eq:h1p}
H_1=\sum_i\frac{1}{2m}\left({\bf p}_i - \frac{e}{c}{\bf A}({\bf
  r}_i)\right)^2 + V({\bf r}_i),
\end{equation}
describing an electron in the magnetic field ${\bf B}={\bf \nabla}
\times {\bf A}$ and confined to the system of quantum dots by the
electrostatic potential $V$.  The quantum dots form in the minima of
this potential, which is locally harmonic with the frequency
$\omega_0$.  The ground states of $H_1$ localized in these wells are
the translated Fock-Darwin states \cite{BLD99}.

The HM Hamiltonian is of the generic form
\begin{equation}
\label{eq:full_h}
\begin{split}
H=&\,  t\, \sum_{k,l}\left(
\delta_{q_k,q_l}\delta_{s_k,s_l}\,c_{q_kIs_k}^{\dagger}c_{q_kOs_k}+{\rm
h.c.}\right) - \\ & \epsilon \sum_{k,p_k=I} c_k^\dagger c_k
+\frac{1}{2}\sum_{klmn}\langle kl|V_C|mn\rangle c_k^\dagger
c_l^{\dagger} c_n c_m.
\end{split}
\end{equation}
The intra-DQD tunneling term $\propto t$ preserves the electron spin.
The bias $\epsilon$ of the inner ($p_k=I$) QDs with respect to outer
($p_k=O$) is taken to be symmetric, i.e., the energy of both inner
dots is lowered by the same amount.  The two-body Coulomb interaction
is denoted by $V_C$.  Near the center of the quantum dot, the
electrostatic potential is approximately harmonic and we assume that
the wave functions of the electrons annihilated by the operators $c_k$
are well approximated by the orthogonalized Fock-Darwin ground states.

The impenetrable barrier that separates the DQDs imposes the
conservation of the number of $L(R)$ electrons, ${\hat
n}_{L(R)}=\sum_{p=I,O;s=\uparrow ,\downarrow}{\hat n}_{L(R)ps}$, where
${\hat n}_{qps}=c_{qps}^{\dagger}c_{qps}$.  The ${\hat n}_{L(R)}$
conserving terms, proportional to the interaction matrix elements
$\langle kl |V_C| mn \rangle$ in Eq.~(\ref{eq:full_h}), where the
indices $k,l,m,n$ denote the single QD ground states, can be divided
into intra-DQD terms where $q_k=q_l=q_m=q_n$ and inter-DQD terms that
satisfy $q_k\neq q_l$ and $q_m\neq q_n$.  All the other terms, e.g.,
the ones that annihilate two electrons on the left ($L$) DQD and
create two on the right ($R$) DQD violate the conservation of the
electron numbers and therefore vanish.

\subsection{Interaction within a double quantum dot}

The terms for the interaction within a DQD in Eq.~(\ref{eq:full_h})
were discussed in \cite{BLD99}.  They renormalize the one-body
tunneling matrix element $t\rightarrow t_H=t+\langle {\bar
S}|V_C|D_{I(O)}\rangle /\sqrt{2}$, introduce the on-site repulsion
$U=\langle D_{I(O)}|V_C|D_{I(O)}\rangle$ of two electrons on the same
QD, and cause transitions between the two doubly occupied DQD states
with the matrix element $X=\langle D_{I(O)}|V_C|D_{O(I)}\rangle$.
Also, the Coulomb interaction on a DQD contributes $V_+=\langle {\bar
S}|V_C|{\bar S}\rangle$ to the electrostatic energy of the symmetric
and $V_-=\langle T_0|V_C|T_0\rangle$ to the antisymmetric singly
occupied orbitals of two electrons in a DQD \cite{BLD99}, giving rise
to a direct exchange interaction between spins.  As a result, the
electrons on a DQD are described by an extended Hubbard model with the
isotropic exchange interaction \cite{BLD99}
\begin{equation}
\label{eq:jbld}
J=V_--V_+-\frac{U_H}{2} +\frac{1}{2}\sqrt{U_H^2+16t_H^2},
\end{equation}
where $U_H=U-V_++X$ is the effective on-site repulsion.

\subsection{Interaction between the double quantum dots}

The Coulomb interaction between the DQDs produces three new classes of
direct terms in the Hamiltonian, while the exchange terms between the
DQD vanish due to the impenetrable barrier.

\begin{figure}[t]
\includegraphics[width=8.5cm]{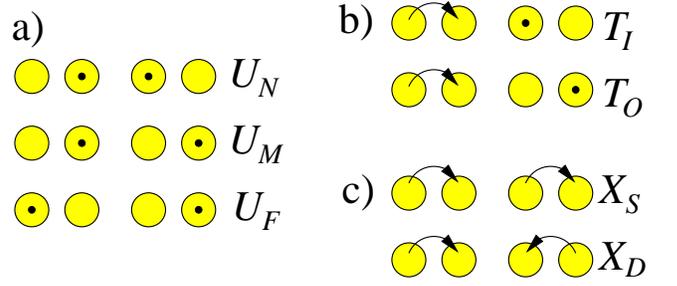}
\caption{ Effects of the direct Coulomb interaction between double
  quantum dots (DQDs).  All the exchange terms between the DQDs vanish
  due to the impenetrable barrier.  a) The Coulomb repulsion between
  the electrons on different double quantum dots contributes to the
  energy of the system.  In the case of identical DQDs separated by an
  impenetrable barrier, there are three such contributions, coming
  from the electrons in orbitals that are near ($U_N$), at a medium
  distance ($U_M$) or far apart ($U_F$).  b) The tunneling matrix
  elements within a DQD are renormalized by $T_I$ or $T_O$, due to the
  interaction with an electron on the inner or the outer dot of the
  other DQD.  c) The interaction enables the correlated hopping
  processes in which electrons simultaneously tunnel in both DQDs.  In
  one such process the electrons tunnel to the same side (either left
  or right) with the matrix element $X_S$.  In the other correlated
  hopping process electrons simultaneously tunnel into the inner or
  outer quantum dots of their double quantum dots with the matrix
  element $X_D$.
\label{fig:correlation}}
\end{figure}

In the first class are the terms proportional to the number operators
${\hat n}_{qps}{\hat n}_{{\bar q} p' s'}$, describing the
electrostatic repulsion of the electrons in states $qps$ and ${\bar
q}p's'$, where ${\bar L}=R$ and ${\bar R}=L$.  For a pair of identical
DQDs, there are three such terms: the interaction  of a pair of
electrons on the inner QDs, $U_N=\langle qIs,{\bar q}Is'|V_C|qIs,{\bar
q}Is'\rangle$, the interaction of an electron on the inner QD of one
DQD and an electron in the outer QD of the other DQD, $U_M=\langle
qIs,{\bar q}Os|V_C|qIs,{\bar q}Os\rangle$, and the interaction of
electrons on the outer QDs, $U_F=\langle qOs,{\bar q}Os|V_C|qOs,{\bar
q}Os\rangle$, Fig.~\ref{fig:correlation}a.

In the second class are the terms proportional to ${\hat
n}_{qps}c^{\dagger}_{{\bar   q}p's'}c_{{\bar  q}{\bar  p}'s'}$, where
${\bar I}=O$ and ${\bar O}=I$.  These terms describe the
spin-independent correction to the tunneling matrix element in the
${\bar q}$ qubit due to the interaction with an electron in the state
$qps$.  The two parameters that determine the tunneling corrections
are $T_{p'}=\langle qps,{\bar q}p's'|V_C|q{\bar p}s,{\bar
q}p's'\rangle$, and are due to the interaction with an electron in the
$p'=I,O$ orbital in the other DQD, Fig.~\ref{fig:correlation}b.

The terms in the third class are proportional to
$c^{\dagger}_{qps}c_{q{\bar p}s}c^{\dagger}_{{\bar q}p's'}c_{{\bar
q}{\bar p}'s'}$, and describe the processes in which electrons in both
DQD tunnel simultaneously, Fig.~\ref{fig:correlation}c.  The two
independent matrix elements for these processes are $X_S=\langle
qps,{\bar q}ps'|V_C|q{\bar p}s,{\bar q}{\bar p}s'\rangle$ describing
the tunneling from the inner to the outer orbital in one DQD and from
the outer to the inner in the other, and $X_D=\langle qps,{\bar
q}{\bar p}s'|V_C|q{\bar p}s, {\bar q}ps'\rangle$ describing the
simultaneous tunneling into inner or outer orbitals in both DQDs.  For
the system in zero magnetic field these two matrix elements are equal,
$X_S=X_D$.

\subsection{Control of the interaction}

In order to describe the influence of the intra-DQD tunneling $t$ and
the bias $\epsilon$ on the spectrum of the DDQD, we have to model the
dependence of the Hamiltonian on these external parameters.  In an
experiment, both $t$ and $\epsilon$ are controlled by applying
voltages to the electrodes that define the quantum dots.  The exact
form of the voltage-dependent DDQD binding potential was studied using
the Schr{\"o}dinger-Poisson equation \cite{WU06}, but here we do not
attempt to calculate the dependence of the Hamiltonian
Eq.~(\ref{eq:hblocks}) on $\epsilon$ and $t$ from first principles.

Instead, we  adopt a quartic double-well model for the potential of a
DQD centered at $(\pm a_2,0)$ of the form \cite{BLD99}
\begin{equation}
\label{eq:doublewell}
V(x,y) = \frac{m\omega_0^2}{2}  \left(\frac{1}{4a_1^2}
\left(\left(x\mp a_2\right)^2-a_1^2\right)^2 + y^2\right),
\end{equation}
where $m$ is the electron effective mass, $2a_1$ is the distance
between the approximately harmonic wells in a DQD, and $2a_2$ is the
distance between the DQD double-well minima.  In the limit of well
separated dots, $a_{1,2}\gg a_{\rm B}$, where $a_{\rm B}$ is the QD
Bohr radius given by $a_{\rm B}^2=\hbar/m\omega_0$, and near the local
minima of the quartic potential well at $(\pm a_2 \pm a_1,0)$, the
potential is approximately harmonic with the frequency $\omega_0$.
The Fock-Darwin ground state wave functions in this harmonic potential
centered at $(x_c,0)$ and in the magnetic field $B$ normal to the
plane of the dots, described in the symmetric gauge by the vector
potential ${\bf A}=B(-y,x,0)/2$, are
\begin{equation}
\label{eq:fd0}
\phi_{x_c}(x,y) = \sqrt{\frac{m\omega}{\pi\hbar}}
  e^{-m\omega\left(\left(x-x_c\right)^2+y^2\right)/2\hbar + im\omega_L
  x_cy /\hbar},
\end{equation}
where $\omega_L=\sqrt{eB/2mc}$ is the electron Larmor frequency and
$\omega=\sqrt{\omega_0^2+\omega_L^2}$ is the resulting confinement
frequency with both electrostatic and magnetic contributions.  We will
use the magnetic compression factor $b=\omega/\omega_0$ to measure the
strength of the magnetic field, consistently with the notation in
\cite{BLD99}.

The translated single-electron Fock-Darwin states $\phi_{\pm a_2\pm
a_1}(x,y)$ define the state space of the variational HM approximation
for a DDQD.  The tunneling matrix element between the Fock-Darwin
ground states in the local minima of the potential
Eq.~(\ref{eq:doublewell}) is our control parameter $t$ \cite{BLD99},
\begin{equation}
\label{eq:todb}
t\equiv \langle \phi_{\pm a_2 +a_1}|H_1|\phi_{\pm a_2 - a_1}\rangle =
\frac{3}{8} \frac{S}{1+S^2} \left(\frac{a_1^2}{a_{\rm
B}^2}+\frac{1}{b}\right),
\end{equation}
where $S=\langle\phi_{\pm a_2 + a_1}|\phi_{\pm a_2
  -a_1}\rangle=\exp(-d_1^2(2b-1/b))$, is the overlap  between the
Fock-Darwin ground states in a DQD.

As $t$ is changed by external voltages, we assume that the overlap $S$
between the oscillator states remains consistent with the relation
Eq.~(\ref{eq:todb}) which is valid for the double-well potential $V$.
All the Coulomb matrix elements can be expressed in terms of $S$ so
that after solving equation (\ref{eq:todb}) for the overlap they
become functions of $t$, see Appendix A.  The bias $\epsilon$ is
modeled as an energy shift of the orbitals, so that the inner $p_k=I$
orbitals have their energy reduced by $\epsilon$.

The two-qubit gates are applied by time-dependent tuning of the
tunneling matrix element $t$ and/or the bias $\epsilon$ in the DQDs
using voltage pulses.  In a typical experiment, the control of the QD
energies through $\epsilon$ is much easier to achieve than the control
over tunneling matrix element $t$ \cite{HB06}.  The reason behind this
is that the energy bias is linear in applied voltage, while the
tunneling is typically exponential.

\begin{figure}[t]
\includegraphics[width=8.5cm]{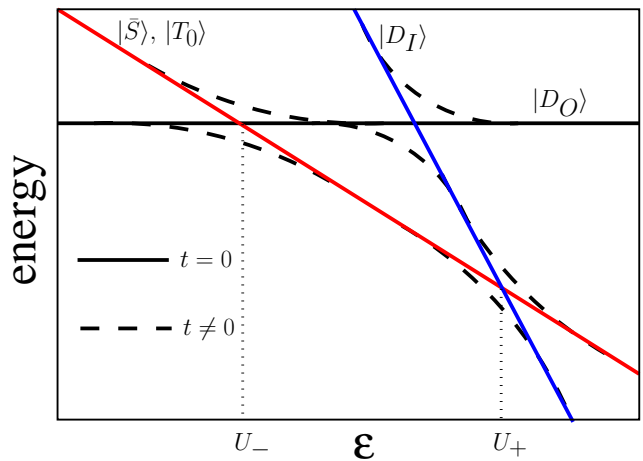}
\caption{Illustration of the double quantum dot energy levels as a
  function of the bias $\epsilon$.  The energy of the singlet state with doubly
  occupied outer quantum dot, $|D_O\rangle$, is independent of the
  bias. The energies of the singly occupied singlet, $|{\bar
  S}\rangle$, and the singly occupied triplet, $|T_0\rangle$, state
  are lowered with the increasing bias as they have a contribution
  $-2\epsilon$ from the biased inner quantum dots.  The energy of the
  singlet with doubly occupied inner quantum dots, $|D_I\rangle$, is
  lowered with the increasing bias faster than the energy of $|{\bar
  S}\rangle$ and $|T_0\rangle$ state, due to the bias contribution of
  $-4\epsilon$.  When the tunneling $t$ is zero, the lowest energy
  levels cross at the bias $U_\pm$, leading to a drastic change of the
  effective spin interaction.  For nonzero tunneling, the levels
  anticross, but the effective spin interaction still changes
  significantly when we tune the system from one side of the
  anticrossing to the other.
\label{fig:flow}}
\end{figure}

\begin{figure}[t]
\includegraphics[width=8.5cm]{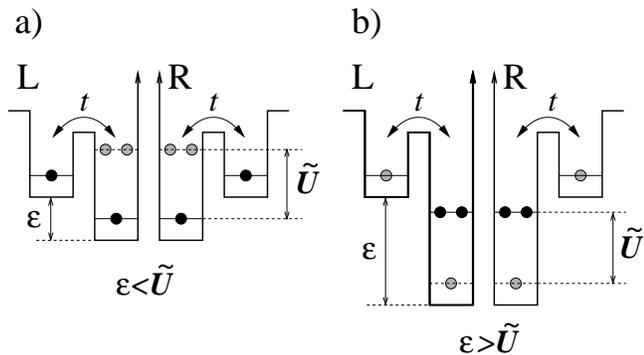}
\caption{Bias dependence of the double-double quantum dot (DDQD)
  ground state.  (a) When the bias $\epsilon$ of the inner quantum
  dots with respect to the outer ones is weaker than the effective
  on-site Coulomb repulsion ${\tilde U}$, the charge configurations of
  the lowest energy singlet and triplet states consists of singly
  occupied orbitals.  (b) When $\epsilon > {\tilde U}$, the lowest
  energy singlet has a doubly occupied inner quantum dot, while the
  orbital state of the lowest energy triplet remains unchanged.
\label{fig:gndbias}}
\end{figure}

The structure of the energy levels is particularly simple in the limit
of zero tunneling $t=0$.  In this limit, the eigenstates are the
Hund-Mulliken basis states,
Eqs.~(\ref{eq:states})--(\ref{eq:statesend}).  Their energies are
determined by the bias $\epsilon$, the external magnetic field $B$,
and the direct Coulomb interaction that is set by the device geometry.
A drastic change in the structure of the DDQD spectrum as a function
of bias $\epsilon$ appears at the crossings of the lowest energy
singlet states within a DQD.  Each of the singlet states $|{\bar
S}\rangle$, $|D_I\rangle$, and $|D_O\rangle$ is lowest in energy for
some values of the bias $\epsilon$, Fig.~\ref{fig:flow}.
A crossing occurs when either the positive bias overcomes the
effective on-site repulsion ${\tilde U}$, making the state with both
electrons in an inner dot $|D_I\rangle$ the lowest in energy, or the
negative bias makes $|D_O\rangle$ the lowest in energy, see
Fig.~\ref{fig:gndbias}.  We use the effective on-site repulsion
${\tilde U}$ to emphasize the fact that it includes not only the
repulsion of two electron in the same dot, denoted by $U$, but also
the energy of the interaction with the electrons on the other DQD.  We
will also use two special values of the effective on-site repulsion,
$U_{\pm}$.  Due to the dependence of the effective on-site repulsion
on the state of the other DQD, the lowest energy singlet-singlet DDQD
state can consist of different singlets on the two dots, as in $|{\bar
S},D_I\rangle$ and $|D_I,{\bar S}\rangle$.  In the strong bias
regions, the lowest energy singlets are doubly occupied states.  For
$\epsilon-U_+ \gg t$ the lowest energy singlet is $|D_ID_I\rangle$,
and for $U_- -\epsilon \gg t$, the lowest energy singlet is
$|D_OD_O\rangle$.  The second doubly occupied singlet state is
separated by an energy gap $\approx |2\epsilon|$ from the lowest
energy state.

\section{Strong Bias}

To develop an intuitive picture of the operation of an entangling
two-qubit gate and the mechanisms for its control, we consider the
simple case of strong bias.  We show how the switching between the
strong bias regime ($\epsilon-U_+ \gg t$), and the weak bias regime in
which the dominant interaction is the on-site repulsion provides us
with control over the entangling interaction $E_e$.  The boundary of
the strong bias regime considered here is set by
$U_+=(3U_N-2U_M-U_F-2V_-+2U)/2$.  A similar strong bias regime with
the lowest energy singlet $|D_O,D_O\rangle$ exists for $U_--\epsilon
\gg t$, where $U_-=(3U_F-2U_M-U_N-2V_-+2U)/2$, but we do not consider
it here in detail.  In both of these regimes, a wide energy gap
$\approx 2|\epsilon|$ to the second doubly occupied state allows us to
neglect that state.  This approximation reduces the dimensions of the
Hamiltonian blocks $H{ab}$, Eq.~(\ref{eq:hblocks}), and allows for a
perturbative solution.

Since the only available DQD states in the strong bias regime are the
triplet $|T_0\rangle$ and two singlets, $|{\bar S}\rangle$ and
$|D_I\rangle$, the $H_{TT}$ block of Eq.~(\ref{eq:hblocks}) is
one-dimensional, $H_{ST}$ and $H_{TS}$ are two-dimensional, and
$H_{SS}$ is four-dimensional.  For the present discussion of the
strong bias regime, we choose the zero of the energy scale at
$4\hbar\omega-2\epsilon+U+2V_++U_N+2U_M+U_F$, setting the expectation
value of the energy of four singly occupied QDs with the DQDs in the
electron singlet states to zero, $\langle {\bar S}, {\bar S}|H|{\bar
S},{\bar S}\rangle=0$.  Using the expressions for the Hamiltonian
matrix elements given in the Appendix A, we find the matrices of the
$H_{ab}$ blocks ($a,b=S,T$).  The energy of the $|TT\rangle$ state is
then
\begin{equation}
E_{TT}=2(V_- -V_+).
\end{equation}
The two-dimensional blocks $H_{TS}$ and $H_{ST}$ are related by the
symmetry under exchange of the double quantum dots $L\leftrightarrow
R$ and in the bases $\lbrace|{\bar S},T_0\rangle, |D_I,T_0\rangle
\rbrace$, and $\lbrace|T_0, {\bar S}\rangle$, $|T_0, D_I\rangle
\rbrace$, have the identical matrix form
\begin{equation}
\label{eq:h_tsst}
H_{TS}=H_{ST}=V_--V_++\left(
\begin{array}{cc}
0 &\sqrt{2}t_S   \\
\sqrt{2}t_S & V_D -\epsilon
\end{array}
\right),
\end{equation}
where $t_S=-t_H+T_S$ is the renormalized hopping matrix element and
$V_D= U - V_+ + U_N - U_F$ is the electrostatic energy cost of doubly
occupying the $p_k=I$ state in the presence of the triplet DQD.  With
our choice of the zero of the energy scale, the ground state energies
of $H_{ST}$ and $H_{TS}$ are
\begin{equation}
\label{eq:jstrong0}
\begin{split}
E_{ST}&=E_{TS} =  \\ &V_- -V_+ +\frac{1}{2}\left(V_D -\epsilon\right)
- \frac{1}{2}\sqrt{(V_D-\epsilon)^2+8t_S^2}.
\end{split}
\end{equation}

From the energies $E_{ST}$ and $E_{TS}$, we extract the isotropic
exchange part of the low-energy four-spin Hamiltonian
Eq.~(\ref{eq:hstrongbias}) as
\begin{equation}
\label{eq:jspectrum}
J=E_{TT}-E_{ST}=E_{TT}-E_{TS}.
\end{equation}

The resulting exchange interaction strength is
\begin{equation}
\label{eq:jstrong}
J = V_- -V_+  -\frac{1}{2}\left(V_D -\epsilon\right) +
\frac{1}{2}\sqrt{(V_D-\epsilon)^2+8t_S^2}.
\end{equation}
Comparing this result with the case of an unbiased isolated double
quantum dot, Eq.~(\ref{eq:jbld}), we see that the effect of the strong
bias $\epsilon$ and the presence of another DQD behind the
impenetrable barrier is the change of the effective on-site repulsion
to the value $V_D - \epsilon$ and a reduction of the effective
tunneling matrix element because of the large gap to the excited doubly
occupied state.  As a consequence of this gap, the isotropic exchange
in the limit of noninteracting DQDs and weak tunneling is $J=V_- - V_+
+ 2 t_H^2/(U -V_+ - \epsilon)$, with the hopping contribution reduced
to a half of the result expected from the standard Hubbard model in
the unbiased case, $4t_H^2/U_H$ \cite{BLD99}.

The four-dimensional block $H_{SS}$ in the basis $\lbrace |{\bar
  S},{\bar S}\rangle,(|{\bar S},D_I\rangle+|D_I,{\bar
  S}\rangle)/\sqrt{2},|D_I,D_I\rangle,(|{\bar S},D_I\rangle-|D_I,{\bar
  S}\rangle)/\sqrt{2},\rbrace$ is
\begin{equation}
\label{eq:h_ss}
H_{SS}=\left(
\begin{array}{cccc}
0 & 2t_S & 2X_D & 0
\\
2t_S & V_D-\epsilon + 2X_S & 2t_I & 0
\\
2X_D & 2t_I & E_{DD} &  0
\\
0 & 0 & 0 & V_D-\epsilon - 2X_S
\end{array}
\right),
\end{equation}
where $t_I$ is the tunneling matrix element renormalized by the
spectator DQD in the doubly occupied state, and
\begin{equation}
\label{eq:edd}
E_{DD}=2U + 3U_N - 2U_M - U_F - 2V_+ - 2\epsilon,
\end{equation}
accounts for the repulsion energy of four electrons in the $p_k=I$
orbitals and the bias $\epsilon$, see Appendix A.  Due to the symmetry
with respect to exchange of the DQDs, $L\leftrightarrow R$, the
antisymmetric state $(|{\bar S},D_I\rangle -|D_I,{\bar
S}\rangle)/\sqrt{2}$ decouples from the other, symmetric, states.

In the limit of large and positive bias, $|\epsilon- V_D|\gg
t_{S/I},X_{S/D}$, all the tunnelling and correlated hopping terms in
the Hamiltonian $H_{SS}$ can be taken to be small.  The unperturbed
Hamiltonian is then diagonal and the ground state energy is $E_{DD}$.
This situation is relevant, because all the small terms are
proportional to the overlap $S$ of the localized states in the quantum
dots, which is small for weakly tunnel-coupled QDs, and we can reach
this regime by applying external voltage to make $|\epsilon- V_D|$
large enough.

Operating the system in the strong bias regime causes a qualitative
change to the effective low-energy Hamiltonian by turning on the
entanglement generating term $E_e$ in Eq.~(\ref{eq:hstrongbias}),
\begin{equation}
\label{eq:ee}
E_e=E_{TT}-2E_{ST}+E_{SS}.
\end{equation}
For weak bias and in the absence of tunneling, the entanglement
generating $E_e$ term is zero, as can be checked from the energies of
the states $|{\bar S},{\bar S}\rangle$, $|T_0,{\bar S}\rangle$, and
$|T_0,T_0\rangle$, given in Appendix A.  This is not true in the case
of strong bias, where the entangling interaction of the strength
$E_{e}=U_N-2U_M+U_F\neq 0$ is present even if the tunneling terms are
zero.  In the strong bias regime, the conditions for $E_e=0$ are
$t_I=t_S$, $X_S=X_D$, and $E_{DD}=2(V_D-\epsilon)$.  While the first
two conditions are satisfied when there is no tunneling, the third is
independent of the tunneling.  It is only satisfied in the limit of
long distance between DQDs, $a_2\gg a_1$, see Fig.~\ref{fig:4dots}.
The tunneling causes a second-order correction to $E_{SS}$,
\begin{equation} \label{eq:ess2}
E_{SS}=E_{DD} + \frac{4t_I^2}{E_{DD}-(V_D - \epsilon)} +
\frac{4X_D^2}{E_{DD}},
\end{equation}
and the corresponding correction to $E_e$ \cite{ft:eeweak}.

We have calculated the matrix elements of the Coulomb interaction
using the basis of single-electron Wannier states obtained by
orthogonalizing the Fock-Darwin ground states centered at the quantum
dots positions, following \cite{BLD99}.  The resulting matrix elements
can all be expressed in terms of the distances between the quantum
dots, and the tunneling matrix element $t$ between QD in DQD.  These
results are summarized in Appendix A.  Together with
Eqs.~(\ref{eq:hstrongbias}), (\ref{eq:jstrong}), and (\ref{eq:ess2}),
they provide a model of the low-energy Hamiltonian of a pair of qubits
realized on a DDQD in the strong bias regime.  This model can describe
a two-qubit quantum gate realized by adiabatically switching the value
of the control parameter $\epsilon$ so that the qubit goes from the
weak bias regime to the strong bias regime and back.

In summary, the interaction of the DQDs causes a change in the
parameters of the extended Hubbard model coupling strength,
Eq.~(\ref{eq:jbld}), so that the energies and hopping matrix elements
on one DQD depend on the state of the other.  Also, the processes in
which the hopping of electrons on the two DQDs is correlated and
mediated by the direct Coulomb interaction become possible, see
Fig.~\ref{fig:correlation}.  The coupling between the DQDs causes an
effective spin interaction that deviates from the form of
exchanged-coupled qubits, adding the entangling term $E_e$ to the
Eq.~(\ref{eq:hstrongbias}).  This deviation creates the entanglement
between the two qubits.  The generation of entanglement can be
efficiently controlled by changing the bias $\epsilon$.

\section{General bias}

The study of a double double quantum dot (DDQD) system in the strong
bias regime presented in Sec.~III allows for a simple perturbative
solution and offers an insight into the mechanism of entanglement
generation.  However, it lacks sufficient predictive power for a
general analysis of a realistic two-qubit quantum gate:  When
switching on and off the entangling interaction, a continuous voltage
pulse is applied, and the system undergoes a smooth transition from
the strong bias regime to the unbiased (or merely biased) regime and
vice versa.  During this transition, the system has to pass through an
intermediate weak-bias regime where the perturbative expansion
Eq.~(\ref{eq:ess2}) breaks down.

In this section, we calculate the full Hund-Mulliken (HM) Hamiltonian
of the four quantum dots, including both $|D_I\rangle$ and
$|D_O\rangle$ states.  This calculation allows us to predict the
quantum gate generated by an arbitrarily shaped adiabatic pulse of the
control parameters $t$ and $\epsilon$.  The main difference in the
system's description is that now we take into account both doubly
occupied states $|D_I\rangle$ and $|D_O\rangle$ in each DQD.
Therefore, we are working in the entire Hilbert space of the HM
approximation, and the strong bias requirement is not important.  Now,
$H_{TT}$ is one-dimensional, $H_{ST}$ and $H_{TS}$ are
three-dimensional, and $H_{SS}$ is nine-dimensional.

Following the discussion of Sec.~III, the effective low-energy spin
Hamiltonian $H$, Eq.~(\ref{eq:hstrongbias}), is determined by the
energies $E_{ab}$, where $a,b=S,T$, of the lowest energy states of a
given spin configuration.  Due to the  $L\leftrightarrow R$ symmetry,
$H$ is the sum of the isotropic exchange terms and the entangling
term.  We proceed by calculating the matrix elements of the
Hamiltonian as a function of the tunneling matrix element $t$ and the
bias $\epsilon$.  The results of this calculation are given in
Appendix A.  Numerical diagonalization of the resulting Hamiltonian
gives the energies $E_{ab}$, for each of the blocks $H_{ab}$, where
$a,b=S,T$. Finally, we extract the effective low-energy Hamiltonian
parameters $J$ and $E_e$ using Eq.~(\ref{eq:jspectrum}) and
Eq.~(\ref{eq:ee}).

\begin{figure}[t]
\includegraphics[width=8.5cm]{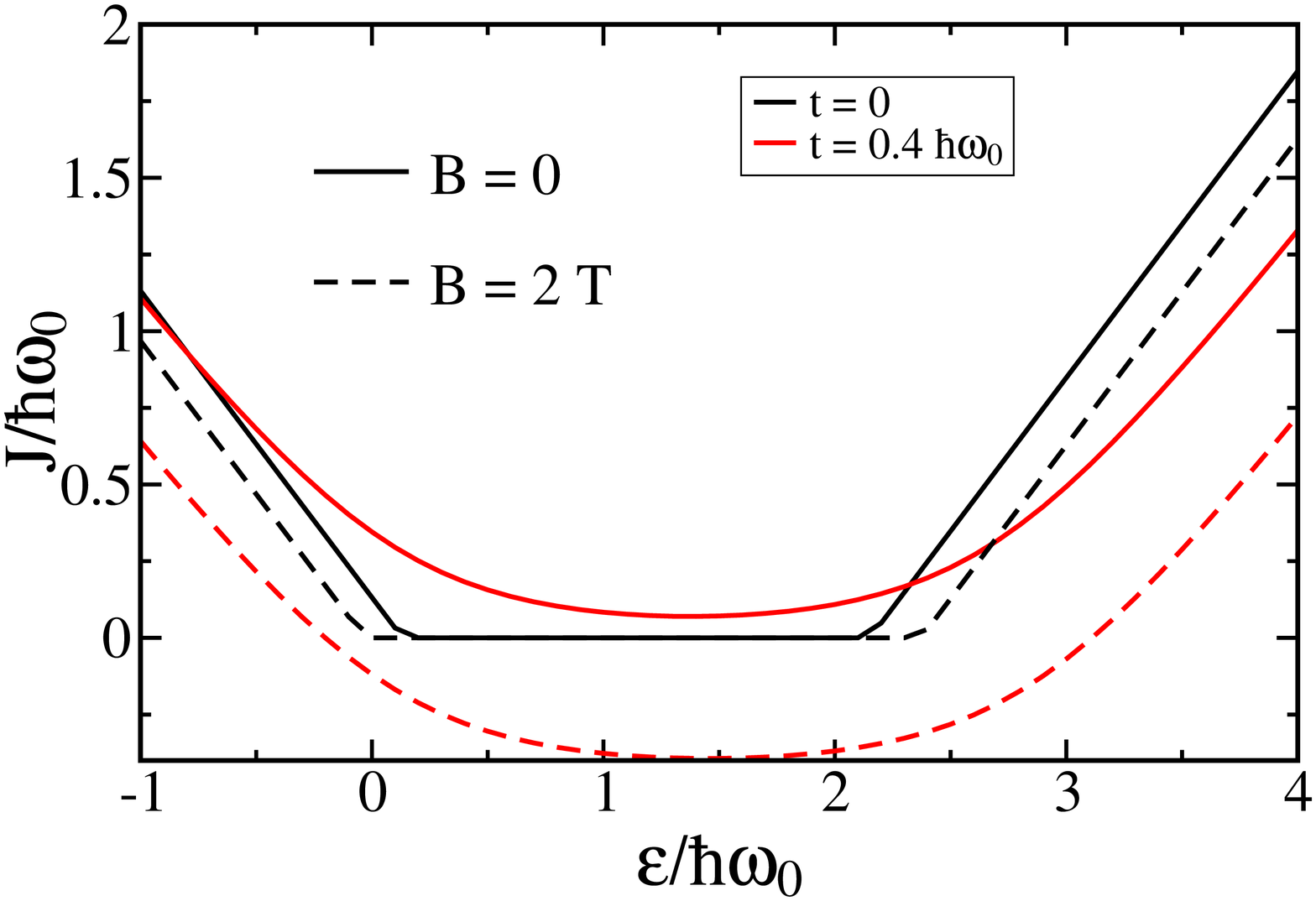}
\caption{Isotropic exchange coupling $J$ as a function of the bias
  $\epsilon$.  In the regions of strong positive and negative bias,
  the exchange coupling is approximately linear $J\propto |\epsilon|$.
  In the intermediate region, the exchange is zero in the zero
  tunneling limit and becomes nonzero as the tunneling is turned on.
  The coupling $J$ is always positive in the absence of a magnetic
  field.  The external magnetic field drives $J$ to negative values in
  a relatively wide range of values of the tunneling matrix element
  and bias.  The confinement energy of the quantum dots is chosen to
  be $\hbar\omega_0=3\,{\rm meV}$, which corresponds to a quantum dot
  Bohr radius $a_{\rm B}=20\,{\rm nm}$ in GaAs.  The distances between
  the dots are chosen to be $2a_{1}=1.6\,a_{\rm B}$ and $2a_2=3\,{\rm
  a_B}$.
\label{fig:jofepsilon}}
\end{figure}

The dependence of the isotropic exchange coupling on the bias
$J(\epsilon)$ is illustrated in Fig.~\ref{fig:jofepsilon}.  In the
zero-tunneling limit, we can identify three regions of qualitatively
different behavior of $J(\epsilon)$.  For strong and negative bias,
$\epsilon < U_-$, corresponding to the $|D_OD_O\rangle$ lowest energy
singlet state, the isotropic exchange coupling is decreasing linearly
with the bias.  In the intermediate region $U_-<\epsilon <U_+$ the
exchange coupling is absent.  For strong and positive bias
$U_+<\epsilon$, the exchange coupling grows linearly with $\epsilon$.
The asymmetric placement of the $J=0$ plateau is a consequence of the
different repulsion energies of the electrons in the inner and outer
QDs.  As the tunneling is turned on, the isotropic exchange couplings
becomes larger due to the mixing of the doubly occupied states in the
plateau region.  For zero magnetic field, the coupling $J$ is
positive.  In a finite field there is a region with negative $J$,
consistent with the analysis of \cite{BLD99} and the experimental
findings of \cite{ZMH+04}.

\begin{figure}[t]
\includegraphics[width=8.5cm]{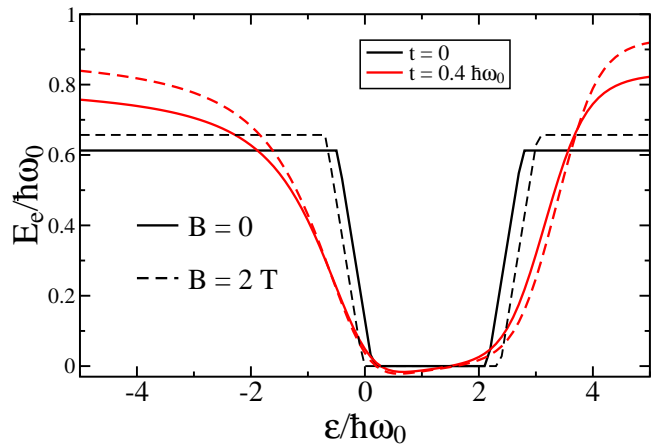}
\caption{Entangling interaction $E_{\rm e}$ as a function of bias.
  The plots correspond to different values of the tunneling matrix
  elements $t$ within the double quantum dots in the absence of a
  magnetic field and in an external magnetic field of $B=2\,{\rm T}$.
  The $t=0$ plot indicates the regions of different lowest energy
  singlets and the positions of crossings.  The strength of the
  entangling interaction $E_{\rm e}$ can be changed significantly by
  tuning the bias $\epsilon$ at a fixed tunneling matrix element $t$.
  Parameters used in this plot are the same as in
  Fig.~\ref{fig:jofepsilon}.
\label{fig:general_bias}}
\end{figure}

A plot of the entanglement generating interaction $E_e$ is given in
Fig.~\ref{fig:general_bias}.
The zero-tunneling value of $E_e$ shows a structure determined by the
Coulomb energies of the basis states
Eqs.~(\ref{eq:states})--(\ref{eq:statesend}).  In a wide plateau of
small bias the entangling interaction vanishes, because all of the
lowest-energy states of definite spin are products of $|{\bar
S}\rangle$ and $|T_0\rangle$.  Since the direct exchange interaction
$V_--V_+$ is zero in the absence of tunneling, those two states are
equal in energy.  When the bias overcomes the on-site repulsion, the
lowest energy states of $H_{SS}$, $H_{ST}$, and $H_{TS}$ change.  The
degenerate lowest energy states of $H_{SS}$ are either $|{\bar S}
D_I\rangle$ and $|D_I {\bar S}\rangle$ in the region of large bias on
the right of the plateau, or $|{\bar S}D_O\rangle$ and $|D_O{\bar
S}\rangle$ in the region of smaller bias to the left of the plateau.
Simultaneously, the analogous states with $|{\bar S}\rangle$ replaced
by $|T_0\rangle$ become the lowest energy states in $H_{ST}$ and
$H_{TS}$.  In these two regions $E_e$ is a linear function of
$\epsilon$, $E_e=U_N-U_F-U-\epsilon$ on the left and $E_e=-U_N+U_F-U
+\epsilon$ on the right of the plateau.  When the absolute value of
the bias is even higher, the lowest energy state in $H_{SS}$ is
$|D_ID_I\rangle$ for a very strong and positive bias and
$|D_OD_O\rangle$ for a very strong and negative bias.  These regions
are characterized by an $\epsilon$-independent $E_e=U_N-2U_M+U_F$ for
large $|\epsilon|$.  The values $U_{\pm}$ for the bias $\epsilon$ at
which the changes in zero-tunneling lowest energy states occur depend
on the geometry of the device, described by the distances $2a_1$ and
$2a_2$ (Fig.~\ref{fig:4dots}) and the quantization energy $\hbar
\omega _0$, and correspond to the changes in behavior of the exchange
coupling strength $J$.

The zero-tunneling case shows a desirable feature in that $E_e$, the
quantity that determines the entanglement between the qubits, can be
switched on and off by tuning $\epsilon$.  However, the regions of
different $E_e$ can not be reached by adiabatic pulses in the
$t\rightarrow 0$ limit.  Turning on the tunneling $t$ between the QDs
will introduce transitions between previously disconnected regions,
and the adiabatic gates become possible.  The simple $t=0$ picture of
the entanglement generated by a difference in Coulomb energies is
perturbed by the transitions.  It is no longer possible to turn off
$E_e$ throughout the plateau region by a change in $\epsilon$ alone.
In the plateau region, $E_e$ is generically nonzero, but small.
Therefore, in order to turn off the entangling interaction when $t$ is
kept constant, it is desirable to keep $t$ small, and to tune
$\epsilon$ to a value where $E_e=0$.

\section{Quantum gate operation}

For a quantum gate applied by the time-dependent Hamiltonian
Eq. (\ref{eq:hblocks}), with the parameters $t$ and $\epsilon$
changing adiabatically on the time scale set by the energy gap between
the states within the blocks $H_{ab}$, the applied gate is determined
by the splittings between the lowest lying states in each of the
subspaces of the definite spin.  If the energies of the lowest energy
states in singlet-singlet, singlet-triplet, triplet-singlet and
triplet-triplet subspaces are $E_{SS}(t)$, $E_{ST}(t)=E_{TS}(t)$, and
$E_{TT}(t)$ respectively, the gate applied by an adiabatic pulse
starting at the time $t_i$ and finishing at $t_f$ will be ${\cal
U}={\rm diag}(\phi_{SS},\phi_{ST},\phi_{TS},\phi_{TT})$, with the
phases
\begin{equation}
\label{eq:phases} 
\phi_{ab}=\exp -\frac{i}{\hbar}\int_{t_i}^{t_f}E_{ab}(t)dt.
\end{equation}

With the ability to turn the entangling interaction on and off and
perform single-qubit gates, it is possible to perform a {\rm CNOT}
gate on a pair of qubits encoded into spin states of DQD.  We consider
a quantum gate implemented by first adiabatically  turning on the
entangling interaction for a period $\tau_{\rm on}$, and then again
adiabatically switching to the Hamiltonian with the entangling
interaction off for the time interval $\tau_{\rm off}$.  The lowest
energy states in each of the $SS$, $ST$, $TS$, and $TT$ subspace will
acquire a phase dependent on the control parameters $\epsilon$ and $t$
and the pulse durations.  In the \textit{on} state, the Hamiltonian
that describes the ground states in all the spin subspaces is, up to a
constant, $H_{\rm on}={\rm diag}(E_e,J_{\rm on},J_{\rm on},2J_{\rm
on})$, where $E_e$ is the strength of the entangling interaction in
the \textit{on} regime, and $J_{\rm on}$ is the corresponding exchange
coupling.  After the DDQD was in the on state for the time $\tau_{\rm
on}$, the applied gate is
\begin{equation}
\label{eq:uon}
{\cal U}_{\rm on}=\exp-i\frac{\tau_{\rm on}}{\hbar}H_{\rm on}.
\end{equation}
Similarly, during the subsequent period of duration $\tau_{\rm off}$
when the entangling interaction is set to zero, the applied gate is
\begin{equation}
\label{eq:uoff}
{\cal U}_{\rm off}=\exp-i\frac{\tau_{\rm off}}{\hbar}H_{\rm off},
\end{equation}
where $H_{\rm off}={\rm diag}(0,J_{\rm off},J_{\rm off},2J_{\rm off})$
in analogy with the \textit{on} regime.  The resulting gate is
\begin{equation}
\label{eq:u}
{\cal U}={\cal U}_{\rm off}{\cal U}_{\rm on}=\exp -i\left(
\begin{array}{cccc}
\phi & 0 & 0 & 0
\\
0 & \lambda & 0 & 0
\\
0 & 0 & \lambda & 0
\\
0 & 0 & 0 & 2\lambda
\end{array}
\right),
\end{equation}
where $\hbar \lambda= J_{\rm on}\tau_{\rm on}+J_{\rm off}\tau_{\rm
off}$ is the integrated strength of the exchange coupling in DQD, and
$\hbar \phi= E_{\rm on}\tau_{\rm on}$ is the integrated strength of
the entangling interaction.

The {\rm CPHASE} gate, which is equivalent to {\rm CNOT} up to single
qubit rotations, is obtained when the gate parameters satisfy
$\phi=m\pi$ and $\lambda=n\pi$, for an odd integer $m$ and an
arbitrary integer $n$.  In order to complete a {\rm CNOT}, we follow a
pulse of on-state Hamiltonian of the duration $\tau_{\rm on}=m\pi
\hbar /E_e$ by a pulse of the off-state Hamiltonian with of the
duration $\tau_{\rm off}=\hbar(n\pi- J_{\rm on}\tau_{\rm
on}/\hbar)/J_{\rm off}$.  The resulting gate is ${\rm
diag}(-1,-1,-1,1)=-{\rm CPHASE}$, for odd $n$ and ${\rm
diag}(-1,1,1,1)$, which is equal to ${\rm CPHASE}$ with the $X$ gate
applied to both qubits before and after ${\cal U}$.  For any integer
$n$,
\begin{equation}
\label{eq:cphase}
{\rm CPHASE}\sim (\xi \otimes \xi){\cal U}(\xi \otimes \xi),
\end{equation}
where $\xi=\exp(i\pi(1+(-1)^n)\sigma _x/4)$.  In order to complete the
{\rm CNOT}, we apply the one-qubit Haddamard gates $H=(X +
Z)/\sqrt{2}$ to the target qubit both before and after the entangling
gate ${\cal U}$. The entire construction can be represented as
\begin{equation}
{\rm CNOT}=({\bf 1}\otimes H)(\xi \otimes \xi){\cal U}(\xi \otimes
\xi)({\bf 1}\otimes H).
\end{equation}
Note that the {\rm CNOT} construction necessarily involves the single
qubit rotations about pseudospin axes different from $z$.  Such
operations can be performed using the asymmetric bias within a DQD
that encodes the qubit in an inhomogeneous external magnetic field
\cite{HB06}.  The entangling part of a {\rm CNOT} gate {\textit can}
be performed by pulsing the bias $\epsilon$ only, and keeping the
tunneling $t$ constant.  Therefore, control over the bias
$\epsilon$ and the availability of an inhomogeneous magnetic field are
sufficient for the universal quantum computing with two-spin qubits.

\section{Conclusion}
 
We have analyzed two-qubit gates in a pair of qubits, each encoded
into singlet and triplet states of a DQD, and coupled by Coulomb
repulsion.  A two-qubit ${\rm CNOT}$ gate, which together with the
single qubit rotations forms a universal set of quantum gates, can be
performed by tuning the bias of the inner dots with respect to the
outer ones.  We identify the entangling interaction strength $E_{\rm
e}$ as a quantity that has to be controlled in order to implement a
${\rm CNOT}$ with the aid of single qubit rotations.

The dependence of $E_{\rm e}$ on the externally controllable bias
$\epsilon$ and the tunneling matrix element $t$ shows that it can in
principle be turned on and off by changing $\epsilon$ alone, if
sufficiently low values of $t$ are available.

The largest change in $E_{\rm e}$ comes from tuning of the system
through the resonance between singly occupied state and doubly
occupied state on a DQD.  At the side of the resonance with a singly
occupied ground state, and far from the resonance, the entangling
interaction $E_e$ is caused by inter-DQD correlation and is small.  On
the other side of the resonance, with a doubly occupied DQD ground
state, the entangling interaction is caused by the direct Coulomb
repulsion and it is much stronger.  Two-qubit gates necessary for a
universal set of gates can be performed by switching between the
strong and weak entanglement generation regimes using voltage pulses.

We thank M. Trif and D. Klauser for discussions.  We acknowledge
funding from the Swiss National Science Foundation (SNF) and through
NCCR Nanoscience.

\appendix

\section{Hund-Mulliken 16 $\times$ 16 Hamiltonian}

The full Hund-Mulliken Hamiltonian is block-diagonal due to the
symmetry of the interactions with respect to arbitrary rotations in
spin space.  In reality, this symmetry is broken by the weak
spin-orbit coupling interaction that we have neglected.  The blocks
are the one-dimensional $H_{TT}$, the two three-dimensional  $H_{TS}$
and $H_{ST}$, and the nine-dimensional $H_{SS}$, where $T$ stands for
a triplet and $S$ for a singlet state on a DQD.  In this Appendix, we
present the matrices of these blocks as functions of the system
geometry and the control parameters.

There is only one $TT$ state and its energy is
\begin{equation}
\label{eq:htt2}
H_{TT}=E_{TT}=2V_- + U_N + 2U_M + U_F - 2\epsilon.
\end{equation}
The three-dimensional blocks $H_{TS}$ and $H_{ST}$ are related by the
symmetry operation of exchanging the DQD and if we choose the basis
$\lbrace |T_0,{\bar S}\rangle,|T_0,D_I\rangle,|T_0,D_O\rangle \rbrace$
for the $TS$ and  $\lbrace |{\bar
S},T_0\rangle,|D_I,T_0\rangle,|D_O,T_0\rangle \rbrace$ for the $ST$
subspace, they can both be represented by the matrix
\begin{equation}
\label{eq:hts4x4}H_{TS}=H_{ST}=\left(
\begin{array}{ccc}
C_{TS} & \sqrt{2}t_S & \sqrt{2}t_S
\\
\sqrt{2}t_S & C_{TI} & X
\\
\sqrt{2}t_S & X & C_{TO}
\end{array}
\right).
\end{equation}
The nine-dimensional block of singlet states, in the direct product
basis composed out of the two-electron states Eq.~(\ref{eq:states}) is
\begin{widetext}
\begin{equation}
\label{eq:hss9x9}
H_{SS}=\left(
\begin{array}{ccccccccc}
 C_{SS} & \sqrt{2}t_S &\sqrt{2}t_S  & \sqrt{2}t_S & 2X_D  & 2X_S &
 \sqrt{2}t_S  & 2X_S & 2X_D
\\
 \sqrt{2}t_S  & C_{SI} & X & 2X_S  & \sqrt{2}t_I  & 0 & 2X_D &
 \sqrt{2}t_I &  0
\\
 \sqrt{2}t_S & X  & C_{SO} & 2X_D  & 0 & \sqrt{2}t_O & 2X_S & 0 &
 \sqrt{2}t_O
\\
 \sqrt{2}t_S  & 2X_S & 2X_D & C_{IS} & \sqrt{2}t_I  & \sqrt{2}t_I & X
 & 0 & 0
\\
 2X_D & \sqrt{2}t_I & 0 & \sqrt{2}t_I & C_{II} & X & 0 & X & 0
\\
2X_S & 0 & \sqrt{2}t_O  & \sqrt{2}t_I  & X  & C_{IO} & 0 & 0 & X
\\
 \sqrt{2}t_S& 2X_D  & 2X_S  & X  & 0 & 0 & C_{OS} & \sqrt{2}t_O &
 \sqrt{2}t_O
\\
 2X_S  & \sqrt{2}t_I  & 0 & 0 & X  & 0 & \sqrt{2}t_O & C_{OI} & X
\\
2X_D  & 0 & \sqrt{2}t_O & 0 & 0 & X  & \sqrt{2}t_O & X  & C_{OO}
\end{array}
\right).
\end{equation}
\end{widetext}
We do not antisymmetrize with respect to the permutations of electrons
that belong to different quantum dots and have non-overlapping orbital
wave functions.  The matrix elements of the Hamiltonian that describe
the Coulomb interaction within a DQD (intra-DQD terms) $U$, $t$, $X$,
$V_+$ and $V_-$ were analyzed in \cite{BLD99}.  The inter-DQD elements
depend on the following matrix elements of the Coulomb interaction
between the product states of the $|qps\rangle$ electrons localized in
the qubit $q$ and the quantum dot $p$, and having a spin $s$,
\begin{eqnarray} 
X_S &=& \langle LIs,RIs'|V_C|LOs,ROs'\rangle,   \\   X_D &=& \langle
LIs,ROs'|V_C|LOs,RIs'\rangle,   \\   T_O &=& \langle
LOs,ROs'|V_C|LIs,ROs'\rangle,      \\      T_I &=& \langle
LIs,RIs'|V_C|LIs,ROs'\rangle.
\end{eqnarray}
In zero magnetic field, we find that $X_S=X_D$.

The off-diagonal elements are determined by
\begin{eqnarray}
t_S &=& T_O + T_I - t_H,
\\
t_I &=& 2T_I - t_H,
\\
t_O &=& 2T_O - t_H,
\end{eqnarray}
and the diagonal elements are given by
\begin{eqnarray}
C_{TT} &=& 2V_- + U_N + 2U_M + U_F - 2\epsilon,
\\
C_{TS} &=& V_+ + V_- + U_N + 2U_M + U_F - 2\epsilon,
\\
C_{TI} &=& V_- + U + 2U_N + 2U_M - 3\epsilon,
\\
C_{TO}&=& V_- + U + 2U_M + 2U_F - \epsilon,
\\
C_{SS} &=& U_N + 2U_M + U_F + 2V_+ -2\epsilon,
\\
C_{SI}&=& 2U_M + 2U_F + U + V_+ - 3\epsilon,
\\
C_{SO} &=& 2U_M + 2U_F + U + V_+ -\epsilon,
\\
C_{II} &=& 4U_N + 2U - 4\epsilon,
\\
C_{IO} &=& 4U_M + 2U -2\epsilon,
\\
C_{OO}&=& 4U_F + 2U,
\end{eqnarray}
where the symmetry with respect to exchange of the DQDs leads to
$C_{AB}=C_{BA}$ where $A,B\in \{T,S,I,O\}$.

To represent the matrix elements in terms of the system parameters,
the single QD quantization energy $\hbar \omega_0$, tunneling matrix
element within an isolated DQD $t$, the bias $\epsilon$ and the
interdot distances $a_1$ and $a_2$, we have to adopt a model for the
binding potential of a DQD and the orbitals of Hund-Mulliken
approximation.  We assume that the QD orbitals are Wannier functions
obtained by orthogonalization of the Fock-Darwin ground states
centered at the positions of the QDs within a DQD, $(a_2 \pm a_1,0)$
and $(-a_2 \pm a_1,0)$.  The Wannier orbitals are of the generic form
\begin{eqnarray}
\label{eq:wannier}
|W_{q,I}\rangle &= N\left(|\phi_{q,I}\rangle -
 g|\phi_{q,O}\rangle\right),
\\
|W_{q,O}\rangle &= N\left(-g|\phi_{q,I}\rangle
 +|\phi_{q,O}\rangle\right),
\end{eqnarray}
where $|\phi_{q,I(O)}\rangle$ is the Fock-Darwin ground state on the
dot belonging to the qubit $q=L,R$ and the inner$(I)$ or outer$(O)$
QD, Eq.~(\ref{eq:fd0}).  The Wannier orbitals are determined by the
overlap of these wave functions,
$S=\langle\phi_{q,I}|\phi_{q,O}\rangle =
\exp\left(-d_1^2\left(2b-1/b\right)\right)$,   through the mixing
$g=(1-\sqrt{1-S^2})/S$ and normalization constant
$N=1/\sqrt{1-2gS+g^2}$.

The Coulomb interaction matrix elements for the DQD centered at $\pm
a_2=\pm d_2 a_{\rm B}$ and QDs within a DQD displaced by $\pm a_1=\pm
d_1 a_{\rm B}$ from the center of the DQD are then expressed as
\begin{widetext}
\begin{eqnarray}
\label{eq:un}
U_N&=&cN^4\left[ \vphantom{\frac{d_1}{2}} f(d_2-d_1,0) +
 2g^2\left(1+S^2\right)f(d_2,0) + g^4f(d_1+d_2,0) + 2S^2g^2f(d_2,d_1)
 - \right.  \\  \nonumber
 &&\left.4gS\left(f(d_2-\frac{d_1}{2},\frac{d_1}{2}) +
 g^2f(d_2+\frac{d_1}{2},\frac{d_1}{2}) \right)
 \vphantom{\frac{d_1}{2}} \right],
\\
\label{eq:uf}
U_F&=&cN^4\left[ \vphantom{\frac{d_1}{2}} f(d_2+d_1,0) +
2g^2\left(1+S^2\right)f(d_2,0) + g^4f(d_2-d_1,0) + 2S^2g^2f(d_2,d_1) -
\right.   \\   \nonumber
&&\left.4gS\left(f(d_2+\frac{d_1}{2},\frac{d_1}{2}) +
g^2f(d_2-\frac{d_1}{2},\frac{d_1}{2})\right) \vphantom{\frac{d_1}{2}} \right],
\\
\label{eq:um}
U_M&=&cN^4\left[ \vphantom{\frac{d_1}{2}} (1+g^4)f(d_2,0) +
g^2\left(f(d_1+d_2,0) + f(d_1-d_2,0) + 2S^2\left(f(d_2,0) +
f(d_2,d_1)\right)\right) - \right.   \\   \nonumber  &&\left.
2gS\left(1+g^2\right) \left(f(d_2+\frac{d_1}{2},\frac{d_1}{2}) +
f(d_2-\frac{d_1}{2},\frac{d_1}{2})\right)  \vphantom{\frac{d_1}{2}} \right],
\\
\label{eq:to}
T_O&=&cN^4\left[ \vphantom{\frac{d_1}{2}}
S\left(\left(1+3g^2\right)f(d_2+\frac{d_1}{2},\frac{d_1}{2}) +
\left(g^4+3g^2\right)f(d_2-\frac{d_1}{2},\frac{d_1}{2})\right) -
\right.   \\   \nonumber
&&\left. \left(g+g^3\right)\left(\left(1+S^2\right)f(d_2,0) +
S^2f(d_2,d_1)\right) - gf(d_2+d_1,0) - g^3f(d_2-d_1,0)
\vphantom{\frac{d_1}{2}} \right],
\\
\label{eq:ti}
T_I&=&cN^4\left[ \vphantom{\frac{d_1}{2}}
S\left(\left(1+3g^2\right)f(d_2-\frac{d_1}{2},\frac{d_1}{2}) +
\left(g^4+3g^2\right)f(d_2+\frac{d_1}{2},\frac{d_1}{2})\right) -
\right.   \\   \nonumber
&&\left.\left(g+g^3\right)\left(\left(1+S^2\right)f(d_2,0) +
S^2f(d_2,d_1)\right) - gf(d_2-d_1,0) - g^3f(d_2+d_1,0)
\vphantom{\frac{d_1}{2}} \right],
\\
\label{eq:xs}
X_S&=&cN^4\left[ \vphantom{\frac{d_1}{2}}
\left(S^2+2g^2+g^4S^2\right)f(d_2,0) + g^2\left(f(d_1+d_2,0) +
f(d_1-d_2,0) + 2S^2f(d_2,d_1)\right) - \right.   \\   \nonumber
&&\left.2S\left(g+g^3\right)\left(f(d_2+\frac{d_1}{2},\frac{d_1}{2}) +
f(d_2-\frac{d_1}{2},\frac{d_1}{2})\right)  \vphantom{\frac{d_1}{2}} \right],
\\
\label{eq:xd}
X_D&=&cN^4\left[ \vphantom{\frac{d_1}{2}}
S^2\left(1+g^4\right)f(d_2,d_1) + g^2\left(f(d_1+d_2,0) + f(d_2-d_1,0)
+ 2\left(1+S^2\right)f(d_2,0)\right) - \right.  \\  \nonumber
&&\left.2S\left(g+g^3\right)\left(f(d_2+\frac{d_1}{2},\frac{d_1}{2}) +
f(d_2-\frac{d_1}{2},\frac{d_1}{2})\right)  \vphantom{\frac{d_1}{2}} \right],
\end{eqnarray}
\end{widetext}
in terms of the overlaps of the harmonic oscillator wave functions
$S$, the mixing factor $g$, and the function
\begin{equation}
f(d,l)=\sqrt{b}\exp\left(-\alpha(d,l)\right)I_0\left(\alpha(d,l)\right),
\end{equation}
where $\alpha(d,l)=bd^2-(b-1/b)l^2$.
We use the contraction factor $b=\omega/\omega_0$ to measure the
 magnetic field strength.  The overall strength of the Coulomb
 interaction is set by $c=\sqrt{\pi/2}e^2/\kappa \hbar \omega _0
 a_{\rm B}$, where $e$ is the electron charge, $\kappa$ is the
 dielectric constant, and $\hbar \omega_0$ is the single isolated QD
 quantization energy \cite{BLD99}.

To model the dependence of the matrix elements on externally
controllable tunneling matrix element $t$, we use the connection
between the tunneling and the overlap $S=S(t)$ that holds for the
quartic double well, Eq.~(\ref{eq:todb}) and assume that it holds
throughout the gate operation.


\begin{thebibliography}{88}
\bibitem{ft:eeweak} We have done a similar calculation of the
second-order correction to the $E_e$ in the weak bias regime, valid in
the region where $U_--\epsilon\gg t \gg \epsilon -U_+$.  The resulting
correction to the $E_e=0$ zeroth-order result contains terms of the
order $X_D^2$ and $X_S^2$.  It is caused by the correlated hopping
processes, and its size is $\propto t^4$.
\end{thebibliography}
\end{document}